\documentclass[final, runningheads]{llncs}
\usepackage[T1]{fontenc}
\usepackage{graphicx}
\usepackage{amsmath}
\usepackage{float}
\usepackage[backend=biber, style=numeric]{biblatex} 
\bibliography{main}
\newcommand{\itadata}{\footnotesize \textsl{ITADATA2024: The 3$^{\text{rd}}$ Italian Conference on Big Data and Data Science}}
\usepackage{fancyhdr}
\fancypagestyle{empty}{%
  \fancyhf{}
  \fancyhead[C]{\itadata}
  \fancyhead[R]{\includegraphics[width=.05\textwidth]{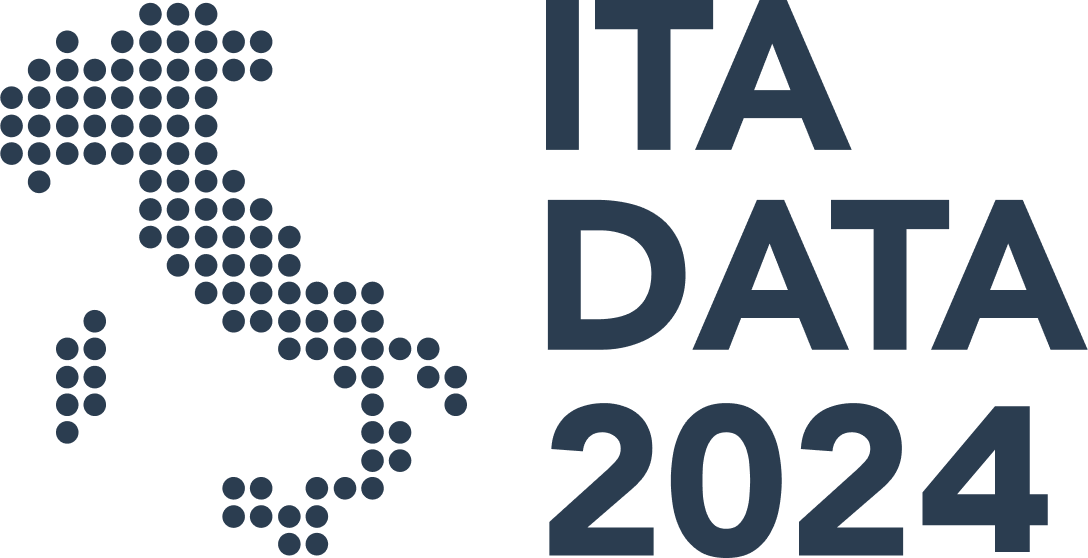}}
}

\usepackage{xcolor}
\usepackage{comment}
\usepackage{subcaption}
\usepackage{booktabs}
\usepackage{cleveref}
\usepackage{xcolor}
\definecolor{greengrass}{rgb}{0.0, 0.7, 0.0}

\begin{document}
\title{A Comparative Analysis of Image Descriptors for Histopathological Classification of Gastric Cancer}
%
%
\author{Marco Usai, Andrea Loddo, Alessandra Perniciano, Maurizio Atzori, Cecilia Di Ruberto}
\authorrunning{Usai et al.}
%
\institute{University of Cagliari, Cagliari, Italy\\
\email{m.usai84@studenti.unica.it, \{andrea.loddo,alessandra.pernician,cecilia.dir,atzori\}@unica.it}}
\maketitle              
\begin{abstract}
Gastric cancer ranks as the fifth most common and fourth most lethal cancer globally, with a dismal 5-year survival rate of approximately 20\%. Despite extensive research on its pathobiology, the prognostic predictability remains inadequate, compounded by pathologists' high workload and potential diagnostic errors. Thus, automated, accurate histopathological diagnosis tools are crucial. This study employs Machine Learning and Deep Learning techniques to classify histopathological images into healthy and cancerous categories. Using handcrafted and deep features with shallow learning classifiers on the GasHisSDB dataset, we offer a comparative analysis and insights into the most robust and high-performing combinations of features and classifiers for distinguishing between normal and abnormal histopathological images without fine-tuning strategies. With the RF classifier, our approach can reach F1 of 93.4\%, demonstrating its validity.

\keywords{Gastric Cancer \and Histopathological Images \and Feature Extraction \and Machine Learning \and Convolutional Neural Networks}
\end{abstract}

\section{Introduction}
Gastric cancer ranks as the fifth most common cancer worldwide and the fourth most lethal, with a global 5-year survival rate of approximately 20\%. Despite extensive research into the disease's pathobiological parameters, predicting disease progression remains challenging, contributing to the low survival rate. Additionally, medical diagnostics' demanding and time-intensive nature can lead to critical details being overlooked during microscopic examinations, potentially resulting in incorrect diagnoses~\cite{ilic2022epidemiology}.

Developing computational tools capable of automatically and accurately performing histopathological diagnoses is crucial to mitigating the issues discussed. Recent advancements in computer technology, particularly in Machine Learning (ML) and Deep Learning (DL), have enabled significant progress in this area. This work explores classifying pathology images into two categories: healthy cells and tumor cells. The study employs various classifiers, extracts specific handcrafted (HC) features, and uses Convolutional Neural Network (CNN) architectures to extract deep features.

This study aims to evaluate the accuracy of shallow learning classifiers in classifying histopathological images using HC and deep features without employing specific fine-tuning strategies. Instead, it relies solely on deep features derived from pre-trained off-the-shelf CNNs.

This work's key contributions include a comparative analysis of various HC and deep features with four ML classifiers and insights into the most robust and high-performing combinations of features and classifiers for distinguishing between normal and abnormal histopathological images.

\section{Materials and Methods}

\subsection{Dataset}
\begin{figure}[t]
    \begin{subfigure}[b]{.4\textwidth}
        \includegraphics[width=\textwidth]{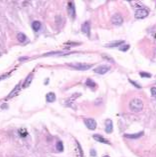}
        \caption{Example of an image belonging to the normal class. It does not contain any cancerous regions.}
        \label{fig:normal}
    \end{subfigure}
    \hfill
    \begin{subfigure}[b]{.4\textwidth}
        \includegraphics[width=\textwidth]{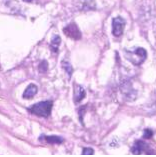}
        \caption{Example of an image belonging to the abnormal class, containing cancerous areas}
        \label{fig:abnormal}
    \end{subfigure}
    \caption{Sample images from the GasHisSDB dataset.}
    \label{fig:gashissdb}
\end{figure}

This work used the GasHisSDB dataset~\cite{kumar2021gashissdb} as the benchmark. It is organized into three subfolders corresponding to different image sizes. This investigation focused on the $160\times160$ subfolder, comprising 33,284 tissue images belonging to the normal (20,160 images) and abnormal (13,124 images) classes. An image is categorized as normal if it lacks cancerous regions, indicating regular microscopic cell observations. Conversely, an image falls into the abnormal class if it contains more than 50\% cancerous areas~\cite{kumar2022comparative}. \Cref{fig:gashissdb} shows two image samples.

\subsection{Image descriptors}\label{sec:features}
HC features extracted from images encompass diverse descriptors aimed at deriving morphological, pixel-level, and textural information. 
As discussed by~\cite{putzu_invariant_2021}, these features are broadly classified into three categories: invariant moments, texture features, and color features. Below, we provide a concise overview of each category and the specific descriptors employed.

\paragraph{Invariant Moments - } An image moment refers to a weighted average, denoted as the moment, of the pixel intensities within an image employed to extract specific properties. Moments find application in image analysis and pattern recognition to characterize segmented objects.

\textbf{Chebyshev Moments (CH)}. These moments are a set of orthogonal moments~\cite{mukundan_image_2001} based on Chebyshev polynomials~\cite{di_ruberto_fast_2018}. In this study, we employed first- and second-order moments, referred to as CH\_1 and CH\_2, respectively. Both were computed of order 5.

\textbf{Second-order Legendre Moments (LM)}. These moments are derived from Legendre orthogonal polynomials~\cite{teh_image_1988,Teague_moments}. They capture and represent objects' shape and spatial characteristics within an image. 
In our analysis, we used Legendre moments of order 5.

\textbf{Zernike Moments (ZM)}. First introduced by~\cite{oujaoura_image_2014}, Zernike moments are a set of orthogonal moments derived from Zernike polynomials. They are utilized to characterize the shape and structure of objects within an image. 
In this study, we applied Zernike moments of order 5 with a repetition of 5.

\paragraph{Texture features - } They are particularly useful to emphasize fine textures.

\noindent\textbf{Rotation-Invariant Haralick Features (HAR)}. 
Thirteen Haralick features~\cite{haralick_textural_1973} were extracted from the Gray Level Co-occurrence Matrix (GLCM) and subsequently converted into rotation-invariant features (refer to~\cite{putzu_rotation_2017} for details). To ensure rotation invariance, four variations of the GLCM were computed, each with a distance parameter $d = 1$ and angular orientations $\theta = [0^\circ, 45^\circ, 90^\circ, 135^\circ]$.

\noindent\textbf{Local Binary Pattern (LBP)}. LBP captures the texture and patterns present in an image, as outlined by~\cite{he_texture_1990}. In our study, we computed the histogram of the LBP, converting it to a rotation-invariant form~\cite{OjalaPM02}, which was then extracted and utilized as the feature vector. The LBP map was generated within a neighborhood defined by a radius $r = 1$ and a number of neighbors $n = 8$.

\paragraph{Color features - } They aim at extracting color intensity information from the images. In this study, these descriptors were calculated from images that underwent a conversion to grayscale, streamlining the process of analysis and computation.

\noindent\textbf{Histogram Features.}
From the histogram that characterizes the overall color distribution within the image, we extracted seven statistical descriptors: mean, standard deviation, smoothness, skewness, kurtosis, uniformity, and entropy. These seven features are collectively referred to as \textit{Hist}.

\noindent\textbf{Autocorrelogram (AC).}
The AC integrates color information with the spatial correlation between colors~\cite{Mitro2016}. Specifically, it records the probability of encountering two pixels of the same color at a distance \(d\). In this study, we utilized four distinct distance values: \(d = 1, 2, 3, 4\). The four resulting probability vectors are concatenated to form a comprehensive feature vector.

\noindent\textbf{Haar Features (Haar)}. These features consist of adjacent rectangles with alternating positive and negative polarities, taking forms like edge features, line features, four-rectangle features, and center-surround features~\cite{viola_rapid_2001}.

\paragraph{Deep features}
In the realm of ML learning, the combination of DL feature extraction and classification alongside shallow learning classifiers has proven to be a potent strategy for enhancing the predictive power of conventional DL models~\cite{bodapati_feature_2019}, also to overcome the obstacle of handling high-dimensional data.

In this context, CNNs have demonstrated their efficacy as feature extractors across various studies~\cite{barbhuiya_cnn_2021, petrovska_deep_2020}. CNNs excel at capturing comprehensive image features by passing input data through multiple convolutional filters, progressively reducing dimensionality across successive architectural layers. For our experiments, we opted for several pre-trained off-the-shelf architectures originating from the Imagenet1k dataset~\cite{deng_imagenet_2009}. Detailed specifications regarding the selected layers for feature extraction, input dimensions, and the count of trainable parameters for each CNN model are outlined in~\Cref{tab:cnn_info}. 
\vspace{-.5cm}

\begin{table*}[!th]
    \centering
    \setlength{\tabcolsep}{6pt}
    \caption{Employed CNN details including number of trainable parameters in millions, input shape, feature extraction layer, and related feature vector size.}
    \label{tab:cnn_info}
    \resizebox{0.95\textwidth}{!}{
    \begin{tabular}{lcclc}
        \toprule
        \textbf{Reference} &\textbf{Parameters (M)} & \textbf{Input shape} & \textbf{Feature layer}& \textbf{\# Features} \\
        \midrule
         AlexNet & 60 & $224\times224$ & Pen. FC & 4,096 \\
         DarkNet-19 & 20.8 & $224\times224$ & Conv19 & 1,000 \\
         DarkNet-53 & 20.8 & $224\times224$ & Conv53 & 1,000 \\
         DenseNet-201 & 25.6 & $224\times224$ & Avg. Pool & 1,920 \\
         EfficientNetB0 & 5.3 & $224\times224$ & Avg. Pool & 1,280 \\
         Inception-v3 & 21.8 & $299\times299$ & Last FC & 1,000 \\
         Inception-ResNet-v2 & 55 & $299\times299$ & Avg. pool & 1,536 \\
         ResNet-18 & 11.7 & $224\times224$ & Pool5 & 512 \\
         ResNet-50 & 26 & $224\times224$ & Avg. Pool & 1,024 \\
         ResNet-101 & 44.6  & $224\times224$ & Pool5 & 1,024 \\
         VGG19 & 144 & $224\times224$ & Pen. FC & 4,096 \\
         XceptionNet & 22.9 & $299\times299$ & Avg. Pool & 2,048 \\
        \bottomrule
    \end{tabular}
    }
\end{table*}

\vspace{-1cm}
\subsection{Machine Learning Algorithms}
\label{sec:mlstrat}
This study's HC and deep features serve as inputs for various ML classifiers. Let us provide a brief introduction to these classifiers.

\textbf{Decision Tree (DT)}. 
A Decision Tree is a hierarchical data structure used for prediction. Each internal node represents a feature, with branches indicating possible feature values. The leaves denote different categories. The algorithm optimizes this structure by pruning nodes that contribute minimally to category separation, thus merging instances at higher levels. Classification is achieved by following the path from the root to a leaf node.

\textbf{k-Nearest Neighbor (kNN)}. 
The kNN classifier categorizes observations by evaluating the classes of the k-training examples closest in distance to the observation in question. This method employs a local strategy to achieve classification by leveraging the proximity of neighboring instances. This work has set $k$ equal to 3.

\textbf{Support Vector Machine (SVM)}. 
SVM differentiates categories by mapping examples to distinct sides of a decision boundary. The one-vs-rest approach is used to handle multiclass problems, training individual classifiers to distinguish each class from all others. Here, a Gaussian radial basis function has been used as the kernel.

\textbf{Random Forest (RF)}. 
This algorithm aggregates predictions from multiple decision trees, each built from random subsets of features and examples. By fostering diversity among the trees, this ensemble method enhances model robustness, improving resilience against data imbalance and mitigating overfitting. This work has set the number of DTs to $100$.

\subsection{Performance Evaluation Measures}
To assess the performance of a binary classifier on a dataset, each instance is categorized as either negative or positive based on the classifier's predictions. The outcome of this classification, compared to the true target value, determines the contribution of each instance to the following measures:
\begin{itemize}
    \item \textbf{True Negatives (TN)}: instances from the negative class correctly predicted as negative.
    \item \textbf{False Positives (FP)}: instances from the negative class incorrectly predicted as positive.
    \item \textbf{False Negatives (FN)}: instances from the positive class incorrectly predicted as negative.
    \item \textbf{True Positives (TP)}: instances from the positive class correctly predicted as positive.
\end{itemize}

Several performance measures have been used to evaluate classification performance. We now provide their definitions, tailored for binary classification tasks.

\begin{itemize}
    \item \textbf{Accuracy (A)}:
    \[
    \text{A} = \frac{TP + TN}{TP + FN + FP + TN}
    \]
    
    \item \textbf{Precision (P)}:
    \[
    \text{P} = \frac{TP}{TP + FP}
    \]
    
    \item \textbf{Recall (R)}:
    \[
    \text{R} = \frac{TP}{TP + FN}
    \]
    
    \item \textbf{Specificity (S)}:
    \[
    \text{S} = \frac{TN}{TN + FP}
    \]
    
    \item \textbf{F1-score (F1)}:
    \[
    F1 = 2 \cdot \frac{\text{P} \cdot \text{R}}{\text{P} + \text{R}}
    \]
    
    \item \textbf{Matthew Correlation Coefficient (MCC)}:
    \[
    MCC = \frac{TP \cdot TN - FP \cdot FN}{\sqrt{(TP + FP)(TP + FN)(TN + FP)(TN + FN)}}
    \]
    
    \item \textbf{Balanced Accuracy (BACC)}:
    \[
    BACC = \frac{\text{R} + \text{S}}{2}
    \]
\end{itemize}

\begin{table}[b]
    \centering
    \caption{Performance obtained with DT trained with HC features.}
    \setlength{\tabcolsep}{6pt}
    \label{tab:hc_dt}
    \begin{tabular}{l c c c c c c c c}
        \toprule
        \textbf{Desc.} & \textbf{A} & \textbf{P} & \textbf{R} & \textbf{S} & \textbf{F1} & \textbf{MCC} & \textbf{BACC} \\
        \toprule
        AC & 62.78 & 69.39 & 68.97 & 53.26 & 69.18 & 22.20 & 61.12 \\
        Haar & 59.70 & 62.56 & 83.31 & 23.43 & 71.46 & 8.33 & 53.37 \\
        Hist & 68.78 & 74.49 & 73.71 & 61.22 & 74.10 & 34.84 & 67.46 \\
        \midrule
        HAR & 68.78 & 74.85 & 72.99 & 62.32 & 73.91 & 35.10 & 67.66 \\
        LBP & 71.22 & 76.23 & 76.26 & 63.47 & 76.25 & 39.74 & 69.87 \\
        \midrule
        CH\_1 & 71.11 & 76.15 & 76.17 & 63.35 & 76.16 & 39.52 & 69.76 \\
        CH\_2 & 71.05 & 76.12 & 76.07 & 63.35 & 76.09 & 39.41 & 69.71 \\
        LM & 71.32 & 76.16 & 76.64 & 63.16 & 76.40 & 39.87 & 69.90 \\
        ZM & 58.06 & 65.74 & 64.24 & 48.57 & 64.98 & 12.73 & 56.40 \\
        \bottomrule
    \end{tabular}
\end{table}

\section{Experimental Results}
In this section, we present the experimental results. The configuration of the experimental setup is detailed in~\Cref{sec:setup}. The outcomes obtained with the HC and deep features are discussed in~\Cref{sec:results_ml} and~\Cref{sec:results_dl}, respectively.

\subsection{Experimental Setup}
\label{sec:setup}

Experiments were conducted on a workstation featuring an Intel(R) Core(TM) i9-8950HK @ 2.90GHz CPU, 32 GB of RAM, and an NVIDIA GTX1050 Ti GPU with 4GB of memory. All implementations and experimental evaluations were executed using MATLAB R2021b.

This study intentionally omitted image augmentation to focus exclusively on extracting pure features from the original images.

The testing strategy employed a 5-fold cross-validation approach. This method ensures statistical reliability by repeatedly training and testing on the same dataset. Specifically, the dataset is divided into 80\% for training and 20\% for testing at each iteration.

\begin{table}[!thbp]
\centering
    \caption{Performance obtained with kNN trained with HC features.}
    \setlength{\tabcolsep}{6pt}
    \label{tab:hc_knn}
    \begin{tabular}{l c c c c c c c c}
        \toprule
        \textbf{Desc.} & \textbf{A} & \textbf{P} & \textbf{R} & \textbf{S} & \textbf{F1} & \textbf{MCC} & \textbf{BACC} \\
        \toprule
        AC & 57.34 & 70.04 & 51.66 & 66.06 & 59.46 & 17.42 & 58.86 \\
        Haar & 42.17 & 56.99 & 18.40 & 78.67 & 27.82 & -3.61 & 48.53 \\
        Hist & 64.64 & 71.97 & 68.15 & 59.24 & 70.01 & 27.07 & 63.70 \\
        \midrule
        HAR & 61.06 & 68.53 & 66.05 & 53.41 & 67.26 & 19.29 & 59.73 \\
        LBP & 69.51 & 75.32 & 73.86 & 62.82 & 74.58 & 36.50 & 68.34 \\
        \midrule
        CH\_1 & 66.16 & 72.41 & 71.28 & 58.29 & 71.84 & 29.45 & 64.78 \\
        CH\_2 & 65.46 & 72.09 & 70.14 & 58.29 & 71.10 & 28.24 & 64.21 \\
        LM & 66.32 & 72.57 & 71.38 & 58.55 & 71.97 & 29.81 & 64.97 \\
        ZM & 57.40 & 65.03 & 64.19 & 46.97 & 64.60 & 11.12 & 55.58 \\
        \bottomrule
    \end{tabular}
\end{table}

\subsection{Results with HC features}
\label{sec:results_ml}

The results of training the classifiers with HC features are shown in~\Cref{tab:hc_dt,tab:hc_knn,tab:hc_rf,tab:hc_svm}. They show the results with DT, kNN, RF, and SVM, respectively.

As shown in~\Cref{tab:hc_dt}, the DT classifier shows accuracy scores from 58.06\% (ZM) to 71.32\% (LM). The LBP and LM features exhibit the best performance, reflecting high precision and BACC, indicating their suitability for DT. Conversely, ZM and Haar features fall short, with lower MCC and specificity values.

kNN (see~\Cref{tab:hc_knn}) presents accuracy scores vary from 42.17\% (Haar) to 69.51\% (LBP). Haar features again exhibit subpar performance, particularly in recall and MCC, suggesting they are unsuitable for kNN. LBP features once more demonstrate superior performance, with high precision and BACC.

Regarding the RF classifier (see~\Cref{tab:hc_rf}), the accuracy scores range from 62.48\% (Haar) to 79.57\% (LBP), indicating a significant variation in the effectiveness of different features. LBP stands out with the highest accuracy and strong precision, recall, and F1 performance. Conversely, Haar features yield poor specificity and MCC, highlighting their inadequacy for this classifier. Interestingly, the CH\_1, CH\_2, and LM features consistently perform well, underscoring their reliability with RF.

Finally, SVM (see~\Cref{tab:hc_svm}) reveals accuracy scores ranging from 44.67\% (Hist) to 75.92\% (CH\_1). CH\_1 and LM features lead with notable precision, recall, and F1-score performance, indicating their robustness for SVM. In contrast, Haar and Hist features are less effective, particularly in recall and specificity, which are crucial for SVM’s discriminative power.

\begin{table}[h]
    \centering
    \caption{Performance obtained with SVM trained with HC features.}
    \setlength{\tabcolsep}{6pt}
    \label{tab:hc_svm}
    \begin{tabular}{l c c c c c c c c}
        \toprule
        \textbf{Desc.} & \textbf{A} & \textbf{P} & \textbf{R} & \textbf{S} & \textbf{F1} & \textbf{MCC} & \textbf{BACC} \\
        \toprule
        AC & 67.30 & 69.18 & 82.99 & 43.20 & 75.45 & 28.71 & 63.09 \\
        Haar & 62.18 & 62.38 & 94.59 & 12.38 & 75.18 & 12.45 & 53.49 \\
        Hist & 44.67 & 96.29 & 9.00 & 99.47 & 16.47 & 17.91 & 54.23 \\
        \midrule
        HAR & 62.07 & 73.64 & 58.21 & 68.00 & 65.02 & 25.64 & 63.10 \\
        LBP & 62.64 & 70.00 & 67.06 & 55.85 & 68.50 & 22.69 & 61.46 \\
        \midrule
        CH\_1 & 75.92 & 77.37 & 85.14 & 61.75 & 81.07 & 48.61 & 73.45 \\
        CH\_2 & 72.50 & 71.57 & 90.55 & 44.76 & 79.95 & 40.78 & 67.66 \\
        LM & 73.32 & 72.65 & 89.73 & 48.11 & 80.29 & 42.61 & 68.92 \\
        ZM & 63.84 & 69.78 & 71.08 & 52.72 & 70.43 & 23.93 & 61.90 \\
        \bottomrule
    \end{tabular}
\end{table}

\begin{table}[h]
    \centering
    \caption{Performance obtained with RF trained with HC features.}
    \label{tab:hc_rf}
    \setlength{\tabcolsep}{6pt}
    \begin{tabular}{l c c c c c c c}
    \toprule
    \textbf{Desc.} & \textbf{A} & \textbf{P} & \textbf{R} & \textbf{S} & \textbf{F1} & \textbf{MCC} & \textbf{BACC} \\
    \toprule
    AC & 71.76 & 72.89 & 84.97 & 51.47 & 78.47 & 39.09 & 68.22 \\
    Haar & 62.48 & 62.65 & 94.22 & 13.71 & 75.26 & 13.61 & 53.97 \\
    Hist & 73.26 & 77.24 & 79.19 & 64.15 & 78.20 & 43.66 & 71.67 \\
    \midrule
    HAR & 76.78 & 78.69 & 84.55 & 64.84 & 81.52 & 50.63 & 74.69 \\
    LBP & 79.57 & 80.74 & 87.03 & 68.11 & 83.77 & 56.61 & 77.57 \\
    \midrule
    CH\_1 & 78.11 & 79.99 & 85.17 & 67.28 & 82.50 & 53.56 & 76.22 \\
    CH\_2 & 78.07 & 79.84 & 85.34 & 66.90 & 82.50 & 53.43 & 76.12 \\
    LM & 78.25 & 80.20 & 85.09 & 67.73 & 82.58 & 53.87 & 76.41 \\
    ZM & 65.19 & 68.16 & 79.84 & 42.70 & 73.54 & 24.26 & 61.27 \\
    \bottomrule
    \end{tabular}
\end{table}

\subsection{Results with deep features} 
\label{sec:results_dl}
The results obtained by training the classifiers with deep features are shown in~\Cref{tab:deep_dt,tab:deep_knn,tab:deep_rf,tab:deep_svm}, with DT, kNN, RF, and SVM, respectively.

DT Classifier with deep features (\Cref{tab:deep_dt}) obtained accuracy scores ranging from 73.52\% (Inception-ResNet-v2) to 84.92\% (DenseNet-201). The latter demonstrates excellent precision, recall, and F1-score, indicating its efficacy. Other deep features, like DarkNet–53 and ResNet-50, also perform well.

kNN with deep features, as shown in~\Cref{tab:deep_knn}, obtained improved results w.r.t. DT since accuracy scores range from 76.13\% (Inception-ResNet-v2) to 88.25\% (DarkNet-53). DarkNet-53 and DenseNet-201 maintain high precision and recall, indicating their robustness. 

The integration of deep features within RF classifier, as detailed in~\Cref{tab:deep_rf}, results in high accuracy scores, ranging from 83.25\% (Inception-ResNet-v2) to 91.93\% (DenseNet-201). DenseNet-201 and EfficientNet B0 features stand out with superior performance across all measures, including MCC and BACC, showcasing their robustness. These results suggest that deep features significantly enhance the RF classifier’s performance.

Finally, the SVM results, shown in~\Cref{tab:deep_svm}, reveal accuracy scores from 69.87\% (Inception-ResNet-v2) to 86.02\% (DenseNet-201). DenseNet-201 and EfficientNet B0 again lead with high precision, recall, and F1.

\vspace{-.5cm}

\begin{table}[!t]
  \caption{Performance obtained with DT trained with deep features.}
  \setlength{\tabcolsep}{6pt}
    \label{tab:deep_dt}%
    \begin{tabular}{l c c c c c c c c}
    \toprule
    \textbf{Desc.} & \textbf{A} & \textbf{P} & \textbf{R} & \textbf{S} & \textbf{F1} & \textbf{MCC} & \textbf{BACC} \\
    \toprule
    AlexNet & 75.00 & 79.47 & 79.19 & 68.57 & 79.33 & 47.72 & 73.88 \\
    DarkNet-19 & 78.25 & 82.04 & 82.04 & 72.42 & 82.04 & 54.46 & 77.23 \\
    DarkNet-53 & 81.64 & 84.78 & 84.95 & 76.57 & 84.86 & 61.55 & 80.76 \\
    DenseNet-201 & 84.92 & 87.51 & 87.60 & 80.80 & 87.56 & 68.42 & 84.20 \\
    EfficientNet B0 & 78.79 & 82.62 & 82.29 & 73.41 & 82.46 & 55.64 & 77.85 \\
    Inception-v3 & 74.54 & 79.21 & 78.60 & 68.30 & 78.90 & 46.81 & 73.45 \\
    Inception-ResNet-v2 & 73.52 & 78.02 & 78.35 & 66.10 & 78.18 & 44.49 & 72.22 \\
    ResNet-18 & 76.73 & 81.20 & 80.13 & 71.50 & 80.66 & 51.46 & 75.82 \\
    ResNet-50 & 81.30 & 84.62 & 84.47 & 76.42 & 84.55 & 60.87 & 80.45 \\
    ResNet-101 & 80.13 & 83.67 & 83.48 & 74.97 & 83.58 & 58.42 & 79.23 \\
    VGG19 & 76.40 & 80.97 & 79.79 & 71.20 & 80.37 & 50.80 & 75.49 \\
    Xception & 79.38 & 83.30 & 82.49 & 74.59 & 82.89 & 56.94 & 78.54 \\
    \bottomrule
    \end{tabular}%
\end{table}

\begin{table}[!bh]
  \centering
    \caption{Performance obtained with kNN trained with deep features.}
    \label{tab:deep_knn}%
    \setlength{\tabcolsep}{6pt}
    \begin{tabular}{l c c c c c c c c}
    \toprule
    \textbf{Desc.} & \textbf{A} & \textbf{P} & \textbf{R} & \textbf{S} & \textbf{F1} & \textbf{MCC} & \textbf{BACC} \\
    \toprule
    AlexNet & 80.85 & 83.80 & 84.77 & 74.82 & 84.28 & 59.78 & 79.80 \\
    DarkNet-19 & 83.99 & 86.47 & 87.20 & 79.05 & 86.84 & 66.40 & 83.13 \\
    DarkNet-53 & 88.25 & 88.89 & 92.11 & 82.32 & 90.48 & 75.25 & 87.22 \\
    DenseNet-201 & 88.21 & 89.04 & 91.84 & 82.63 & 90.42 & 75.16 & 87.23 \\
    EfficientNet B0 & 87.53 & 89.01 & 90.60 & 82.82 & 89.80 & 73.79 & 86.71 \\
    Inception-ResNet-v2 & 76.13 & 79.89 & 80.98 & 68.69 & 80.43 & 49.85 & 74.83 \\
    Inception-v3 & 80.88 & 83.65 & 85.04 & 74.48 & 84.34 & 59.80 & 79.76 \\
    ResNet-101 & 87.89 & 88.57 & 91.87 & 81.79 & 90.19 & 74.48 & 86.83 \\
    ResNet-18 & 84.21 & 85.89 & 88.47 & 77.68 & 87.16 & 66.73 & 83.07 \\
    ResNet-50 & 86.97 & 88.20 & 91.09 & 80.41 & 89.62 & 73.19 & 85.75 \\
    VGG19 & 83.35 & 85.94 & 86.94 & 77.95 & 86.44 & 65.88 & 82.45 \\
    Xception & 85.15 & 86.74 & 89.44 & 78.95 & 88.07 & 70.17 & 84.20 \\
    \bottomrule
    \end{tabular}%
\end{table}

\begin{table}[h]
  \centering
  \caption{Performance obtained with RF trained with deep features.}
  \setlength{\tabcolsep}{6pt}
  \label{tab:deep_rf}%
    \begin{tabular}{l c c c c c c c c}
    \toprule
    \textbf{Desc.} & \textbf{A} & \textbf{P} & \textbf{R} & \textbf{S} & \textbf{F1} & \textbf{MCC} & \textbf{BACC} \\
    \toprule
    AlexNet & 84.02 & 85.55 & 88.57 & 77.03 & 87.03 & 66.29 & 82.80 \\
    DarkNet-19 & 88.30 & 88.68 & 92.49 & 81.87 & 90.54 & 75.33 & 87.18 \\
    DarkNet-53 & 90.30 & 90.72 & 93.55 & 85.30 & 92.11 & 79.58 & 89.42 \\
    DenseNet-201 & 91.93 & 92.61 & 94.20 & 88.46 & 93.40 & 83.05 & 91.33 \\
    EfficientNet B0 & 89.89 & 89.96 & 93.77 & 83.92 & 91.83 & 78.71 & 88.85 \\
    Inception-v3 & 85.52 & 85.64 & 91.42 & 76.46 & 88.44 & 69.39 & 83.94 \\
    Inception-ResNet-v2 & 83.25 & 84.10 & 89.21 & 74.10 & 86.58 & 64.55 & 81.65 \\
    ResNet-18 & 86.99 & 87.32 & 91.87 & 79.50 & 89.53 & 72.54 & 85.68 \\
    ResNet-50 & 89.92 & 90.12 & 93.63 & 84.23 & 91.84 & 78.77 & 88.93 \\
    ResNet-101 & 89.59 & 89.76 & 93.48 & 83.62 & 91.58 & 78.07 & 88.55 \\
    VGG19 & 85.98 & 86.61 & 90.92 & 78.40 & 88.71 & 70.41 & 84.66 \\
    Xception & 88.58 & 89.08 & 92.49 & 82.59 & 90.75 & 75.94 & 87.54 \\
    \bottomrule
    \end{tabular}%
\end{table}

\begin{table}[!ht]
    \centering
    \caption{Performance obtained with SVM trained with deep features.}
    \setlength{\tabcolsep}{6pt}
    \label{tab:deep_svm}
    \begin{tabular}{l c c c c c c c c}
    \toprule
    \textbf{Desc.} & \textbf{A} & \textbf{P} & \textbf{R} & \textbf{S} & \textbf{F1} & \textbf{MCC} & \textbf{BACC} \\
    \toprule
    AlexNet & 72.97 & 77.49 & 76.56 & 65.81 & 77.02 & 42.80 & 71.18 \\
    DarkNet-19 & 77.86 & 82.20 & 80.82 & 71.95 & 81.50 & 53.28 & 76.39 \\
    DarkNet-53 & 82.83 & 85.84 & 84.99 & 76.88 & 85.41 & 63.65 & 80.94 \\
    DenseNet-201 & 86.02 & 89.01 & 86.61 & 81.84 & 87.79 & 69.91 & 84.23 \\
    EfficientNet B0 & 82.84 & 85.76 & 85.35 & 76.96 & 85.55 & 63.48 & 81.15 \\
    Inception-v3 & 73.60 & 78.88 & 74.73 & 67.51 & 76.75 & 45.86 & 71.12 \\
    Inception-ResNet-v2 & 69.87 & 75.03 & 71.65 & 62.78 & 73.29 & 35.68 & 67.21 \\
    ResNet-18 & 77.78 & 82.17 & 80.82 & 71.86 & 81.49 & 53.13 & 76.34 \\
    ResNet-50 & 82.77 & 86.11 & 84.34 & 77.87 & 85.22 & 63.43 & 81.11 \\
    ResNet-101 & 82.52 & 85.76 & 84.71 & 76.81 & 85.23 & 63.00 & 80.76 \\
    VGG19 & 79.60 & 83.94 & 81.78 & 73.70 & 82.84 & 57.67 & 77.74 \\
    XceptionNet & 82.24 & 85.75 & 84.22 & 76.22 & 84.97 & 62.53 & 80.22 \\
    \bottomrule
    \end{tabular}%
\end{table}

\subsection{Discussion}
Across all classifiers, DenseNet-201 and EfficientNet B0 features consistently deliver superior performance. The RF and kNN classifiers benefit from deep features, achieving high accuracy and balanced performance. Conversely, traditional HC features, while sometimes effective, generally lag behind deep features, emphasizing the shift towards DL-based feature extraction in modern ML workflows. This comprehensive evaluation underscores the importance of selecting appropriate feature sets tailored to specific classifiers to optimize performance outcomes. Also, it demonstrates that the task at hand can be faced with general-purpose features without specific fine-tuning strategies.

Compared to the reference results provided by the dataset's authors, we achieved results comparable to those obtained with end-to-end fine-tuned DL methods in this work. 
For example, the RF classifier reached a maximum F1 of 93.4\%, being only 3\% lower than the best result obtained in~\cite{kumar2021gashissdb} with a fine-tuned ResNet-50.

\section{Conclusion}
In this work, the task of histopathological image classification has been faced with a comprehensive analysis of several different HC and deep features used to train four different ML classifiers. We have shown that deep features, like those extracted from DenseNet-201 and EfficientNet B0, obtained superior performance than HC, demonstrating the power of advanced CNN architectures in extracting relevant features for classification tasks. An RF classifier reached a maximum F1 of 93.4\%, highlighting the efficacy of our approach in achieving competitive results without fine-tuning more complex systems.

Despite these promising results, future work will extend classification to include feature merging and further classifiers, incorporating Vision Transformers. Additionally, further developments will address cross-dataset issues through testing and potential refinements on new datasets and different types of staining.

\vspace{3cm}
\printbibliography

\end{document}